**RESEARCH**

**Open Access**

# Evidence-based hand hygiene: Liquid or gel handrub, does it matter?

Constantinos Voniatis[1,2], Száva Bánsághi[3,6], Dániel Sándor Veres[4], Péter Szerémy[5], Angela Jedlovszky-Hajdu[2], Attila Szijártó[1] and Tamás Haidegger[3,5,6*]

## Abstract

**Background**  Recent studies put under scrutiny the prevailing hand hygiene guidelines, which incorporate quantitative parameters regarding handrub volume and hand size. Understanding the criticality of complete (i.e., efficient) hand hygiene in healthcare, objectivization of hand hygiene related parameters are paramount, including the formulation of the ABHR. Complete coverage can be achieved with optimal Alcohol-Based Hand Rub (ABHR) provided. The literature is limited regarding ABHR formulation variances to antimicrobial efficiency and healthcare workers' preference, while public data on clinically relevant typical application differences is not available. This study was designed and performed to compare gel and liquid format ABHRs (the two most popular types in Europe) by measuring several parameters, including application time, spillage and coverage.

**Methodology**  Senior medical students were invited, and randomly assigned to receive pre-determined ABHR volumes (1.5 or 3 ml). All the 340 participants were given equal amounts of gel and liquid on two separate hand hygiene occasions, which occurred two weeks apart. During the hand hygiene events, by employing a digital, fully automated system paired with fluorescent-traced ABHRs, disinfectant hand coverage was objectively investigated. Furthermore, hand coverage in relation to the participants' hand sizes was also calculated. Additional data collection was performed regarding volume differences and their effect on application time, participants' volume awareness (consciousness) and disinfectant spillage during the hand hygiene events.

**Results**  The 1.5 ml ABHR volume (commonly applied in healthcare settings) is insufficient in either formulation, as the non-covered areas exceeded significant (5%+) of the total hand surface area. 3 ml, on the contrary, resulted in almost complete coverage (uncovered areas remained below 1.5%). Participants typically underestimated the volume which they needed to apply. While the liquid ABHR spreads better in the lower, 1.5 ml volume compared to the gel, the latter was easier handled at larger volume. Drying times were 30/32 s (gel and liquid formats, respectively) when 1.5 ml handrub was applied, and 40/42 s when 3 ml was used. As the evaporation rates of the ABHR used in the study are similar to those available on the market, one can presume that the results presented in the study apply for most WHO conform ABHRs.

**Conclusion**  The results show that applying 1.5 ml volume was insufficient, as large part of the hand surface remained uncovered (7.0 ± 0.7% and 5.8 ± 1.0% of the hand surface in the case of gel and liquid, respectively) When 3 ml handrub was applied drying times were 40 and 42 s (gel and liquid, respectively), which is a very long time in daily clinical practice. It looks like we cannot find a volume that fits for everyone. Personalized, hand size based ABHR volumes may be the solution to find an optimal balance between maximize coverage and minimise spillage and drying time. 3 ml can be a good volume for those who have medium size hands. Large handed people should use more handrub to reach appropriate coverage, while small-handed ones may apply less to avoid massive spillage and not to take unrealistically long to dry.

*Correspondence:
Tamás Haidegger
haidegger@irob.uni-obuda.hu
Full list of author information is available at the end of the article

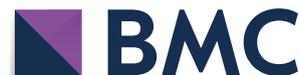

## Background

For decades, hand hygiene protocols and guidelines have been published, re-evaluated, modified and re-written [1–3], however, changes and improvements are still necessary due to the importance of the field. In the past few years, not only have we documented the rise of highly resistant strains [4], but also a pandemic took tolls [5, 6]. Alcohol-Based Handrubs (ABHRs) are universally utilised in hospitals and healthcare institutions worldwide, and while some attempts have been made advocating non-alcohol based handrubs [7, 8], researchers and physicians both agree that ABHRs are the most efficient and effective products preventing contact infection transmission. Nevertheless, numerous blank spots still exist regarding the science of their actual application.

While current hand hygiene research is predominantly focused on issues, like enhancing the antimicrobial effect of disinfectants [4, 9], or increasing the hand hygiene compliance of healthcare-workers (HCW) [7, 10, 11], important questions of application still remain unanswered. The most widely followed 2009 World Health Organization guideline regarding patient safety and hand hygiene is accurate and comprehensive with respect to the workflow points and timing of hand hygiene events. However, specifications for crucial parameters, like ABHR volume are overlooked, the only instructing being the application of a 'palmful' of ABHR [1], while gel and foam formulations are not even mentioned. Similarly, CDC guidelines do not define required ABHR quantity, only mention that the entire hand surface should be covered [2, 12].

Recently, more studies investigated whether the addition of a specified volume or hand size-dependent volume would improve efficiency [13–15]. Our previous results demonstrated how complete hand coverage highly depends on hand size [16–18], and that the currently prescribed 20–30 s application time is unrealistic, as complete hand surface coverage requires longer application times [13, 14, 18], even if we take into account that different alcohol concentrations of ABHR can affect the drying time [19].

In the same way, the formulation (a.k.a. format) of the ABHR (i.e., liquid, gel or foam) is also a crucial, and undervalued point, not incorporated in current guidelines. Even more, according to our knowledge, no actual data can be found regarding the practical and clinically relevant application differences regarding application time, spillage and hand surface coverage.

A few investigations documented the compliance rates of different ABHR formats [19, 20]. In a nutshell, HCWs typically prefer one to the other based on their personal experience, and not following evidence-based research [20–24], nonetheless, such currently is almost non-existent. Our present-day knowledge concludes that the antibacterial efficiency of the gels and liquid ABHRs is equal, at least in an in vitro setting [25, 26]. Due to their physical properties, liquids can spread faster, yet they are prone to spillage (causing material waste). On the contrary, gels are having higher viscosity, aiming to increase coverage, while preventing spillage, but seem to require longer application times and tend to cause sticky hands (decreasing overall compliance) [20]. According to our investigations, gel-based ABHRs cost more than liquid ones, most probably due to their additional components, e.g., hydroxy methylcellulose and glycerine, making the choice of ABHR formulation complex.

he current study aimed to compare the real-life clinical applicability of liquid and gel format ABHRs by objectively measuring several influencing parameters, including hand coverage by handrub, application time, and spillage with the addition of measuring participants hand sizes and in vitro evaporation rate of different products.

## Methods

### Participants

All participants in this study were medical students (N = 340) of the Semmelweis University (Budapest, Hungary). All students who attend class at our clinic were invited to participate in the study, voluntarily. Students undertook a short course on hand hygiene, after which they received individual RFID cards to record their individual data (Table 1).

### Experimental settings

Measurements involved giving participants pre-determined, randomly assigned, exact volumes (1.5 or 3 ml) of either liquid or gel disinfectant (Table 2). A Dispensette S Analog-adjustable bottle-top dispenser (Brand GmbH, Germany) was used to apply the liquid, while a

**Table 1** ABHR included in the evaporation rate measurement

| Manufacturer | Product | Format | Active ingredients (w/w %) |
| --- | --- | --- | --- |
| ANTISEPTICA Dr. Hans Joachim Molitor GmbH | Manorapid | Liquid | 63% isopropanol, 14% propan-1-ol |
| ANTISEPTICA Dr. Hans Joachim Molitor GmbH | Poly-Alcohol Hände-Antisepticum | Liquid | 70% isopropanol |
| B. Braun Medical AG | Promanum Pure | Liquid | 73% ethanol, 10% isopropanol |
| B. Braun Medical AG | Softa-Man | Liquid | 45% ethanol, 18% propan-1-ol |
| BIOETANOL AEG Sp. z o. o | Bioseptol 80 | Liquid | 80–85% ethanol, 10% isopropanol |
| BODE Chemie GmbH | Manusept Basic | Liquid | 80% ethanol |
| BODE Chemie GmbH | Sterillium | Liquid | 45% isopropanol, 30% propan-1-ol, 0.2% mecetroniumetilsulfate |
| BODE Chemie GmbH | Sterillium Classic Pure | Liquid | 45% isopropanol, 30% propan-1-ol, 0.2% mecetroniumetilsulfate |
| Lysoform Schweizerische Gesellschaft für Antisepsie AG | Aco-Derm V | Liquid | 75% ethanol |
| MOLAR chemicals LTD | Semmelweis Training Rub | Liquid | 70% ethanol |
| Schülke & Mayr GmbH | Desderman Pure | Liquid | 75% ethanol, 0.1% biphenyl-2-ol |
| Schülke & Mayr GmbH | Desmanol Pure | Liquid | 75% isopropanol |
| UNICLEAN Kft | Clarasept Derm | Liquid | 52% ethanol, 6.6% bifenil-2-ol |
| B. Braun Medical AG | Softa-Man ViscoRub | Gel | 45% ethanol, 18% propan-1-ol |
| BIOETANOL AEG Sp. z o. o | Bioseptol Gel | Gel | |
| BODE Chemie GmbH | Sterillium Gel | Gel | 85% ethanol |
| Florin Zrt | Bradolife Gel | Gel | 73% ethanol |
| JohnsonDiversey UK Limited | Soft Care Med H5 | Gel | 50–75% isopropanol |
| MOLAR Chemicals LTD | Semmelweis Training Gel | Gel | 70% ethanol |

**Table 2** Gel versus Liquid Investigation Arrangements

| | 1st measurement | 2nd measurement | No. of participants |
| --- | --- | --- | --- |
| Group #1 | 1.5 ml gel | 1.5 ml liquid | 99 |
| Group #2 | 1.5 liquid | 1.5 ml gel | 61 |
| Group #3 | 3 ml gel | 3 ml liquid | 80 |
| Group #4 | 3 ml liquid | 3 ml gel | 71 |

Σ = 311, as 29 students did not participate in the second event calibrated Purell ADX-7 (GOJO Industries Inc., Akron, OH) dispenser was used for the gel. Disinfectants were placed on the centre of the dominant hand's palm. All measurements were performed under the direct supervision of a qualified investigator. Participants did not get any instruction on how they should rub their hands, they were only asked to reach complete coverage. Students were previously trained to follow the WHO 6-step protocol [1].

### In vitro evaporation rate investigation

During the study, we compared two commercially available ABHR products. In order to exclude any possible bias, and to prove the findings can be generalized, regardless of ABHR brand, an in vitro evaporation rate investigation of commercially available products was conducted. Evaporation rates were examined on two typically applied volumes (1.5 and 3 ml) on three different sized Petri dishes (surface areas: 7.07, 19.63 and 38.48 cm$^2$). To assess the evaporation rates, disinfectants were first taken to Petri dishes. Then, to replicate the hand temperature, the covered Petri dishes were placed on a heater, adjusted to 34 °C [27, 28]. The mass of the disinfectant was measured when the cover was removed, and every 30 s after, for up to 3 min. Evaporation was calculated by subtracting the mass registered every 30 s from the initial mass at t = 0. Measurements occurred under standard ambient laboratory conditions (25 ± 3 °C, 30 ± 5% humidity) and were repeated 5 times with each handrub (Fig. 1). The list of the investigated commercial handrubs can be found in Table 1.

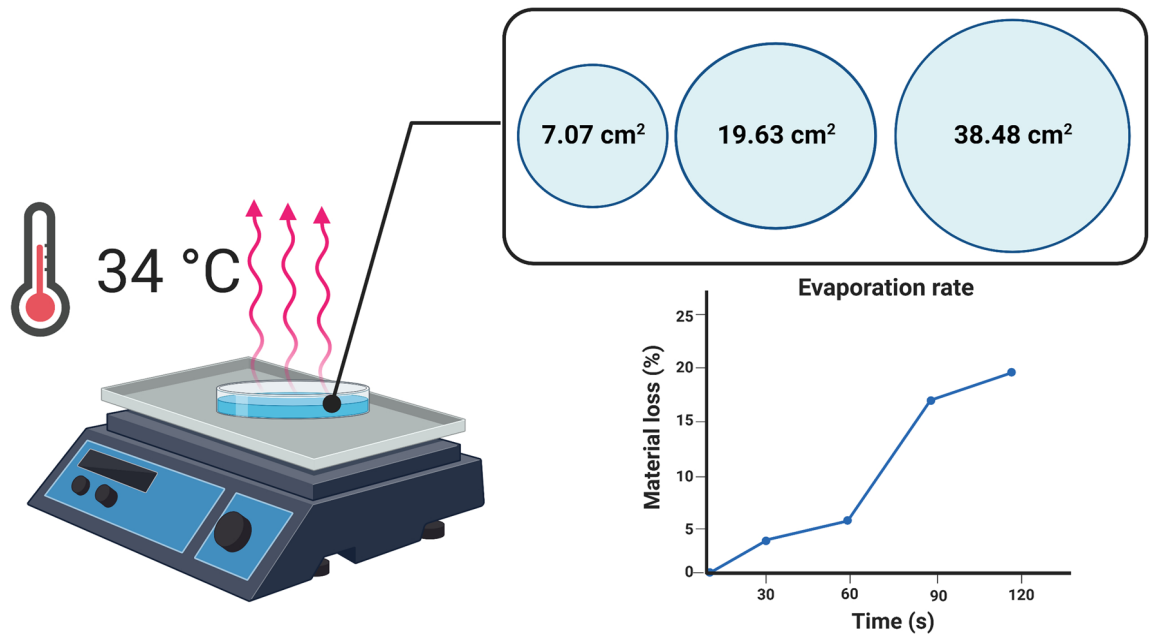

**Fig. 1** Evaporation rate measurement setup

## Hand coverage measurements

Distribution, and consequently hand coverage differences between liquid and gel formulated ABHRs were measured employing an innovative electronic and completely automated digital health technology system (Semmelweis Hand Hygiene System, HandInScan Zrt., Debrecen, Hungary), which has been shown to be superior to any human expert-based evaluation method [29]. Furthermore, application times, volume awareness and disinfectant spillage were also assessed for every single hand hygiene event.

The Semmelweis Hand Hygiene System (Fig. 2) can evaluate the disinfectant's coverage on the hand. By adding a fluorescent dye to the handrub, whether it is a liquid

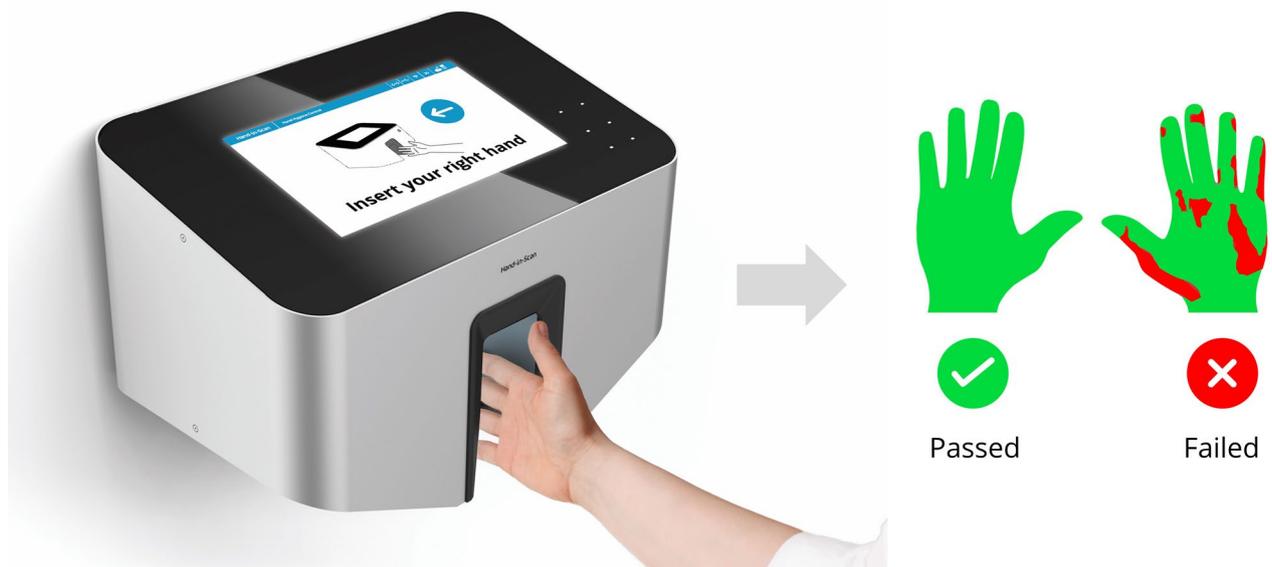

**Fig. 2** Measurements using the Semmelweis System, where green indicates handrub-covered areas, red indicates non-covered areas determined by an AI algorithm, based on image analysis performed on recorded hands, treated with an UV-dyed alcohol-based solution. (Image credit: HandInScan Zrt)

or gel based one, the device can detect the covered (and validly disinfected) areas with a pixel sized resolution. The system presents an immediate and quantitative feedback for health care workers to examine their hand rubbings' efficiency.

Both the gel and the liquid (Semmelweis Training Rub, Semmelweis Training Gel, Molar Chemicals Kft.) had a similar alcohol concentration (70% ethanol), and contained a small amount of fluorescent dye (< 0.02%). After the hand rubbing, participants' hands were assessed employing the Semmelweis System.

### Hand size determination

Hand size for every participant was calculated by counting the pixels of the systems' scanned images. Poor quality images were excluded from the evaluation. Hand surface ($cm^2$) assessment was performed according to an already established *Automated Area Assessment* method [17], which estimates hand size by using a calibration to convert pixel counts to centimetres, and thus considers that the 3D surface of the hands is 1.36× bigger as their two dimensional projections.

### Application time measurements

Application time was measured by a digital stopwatch. Participants were given a clear signal when to start the hand rubbing. As soon as participants felt that their hands were dry, they signalled to the investigator, who recorded and documented the exact time.

### Volume awareness

Adjacent to the physical experiments, participants were surveyed on how the given disinfectant volume felt (Subjective Volume Awareness Assessment, presented in Table 3), before the hand rubbing's evaluation with the Semmelweis System. The participants were not informed about the exact volume they were given during the experiments.

**Table 3** Subjective volume awareness assessment

| Question | Possible answers |
|---|---|
| How did you find the disinfectant volume? | Not enough/just right/too much |

### Disinfectant spillage

To assess the disinfectant spillage, a custom experiment design including an A3-size paper was used. Participants were required to perform hand rubbing directly above the sheet of paper (66,329,742 celeste chiaro coloured paper, A3-size, 80 $g/m^2$ grammage, Fabriano, Italy) at approximately 20 cm height. After an extensive drying period, papers were photographed using a camera equipped with a UV flashlight. The disinfectant droplets that stained the paper were fluorescent, therefore could be assessed using suitable software evaluation (segmentation with custom written segmentation algorithm in Python). Utilizing control measurements as a reference, the spillage was evaluated according to the fluorescence patches' overall intensity and area. Additionally, students were asked whether they felt the disinfectant dripping from their hands during the hand rubbing process (Table 4). The setup was built so that participants could not directly verify the spillage with their own eyes.

### Statistical analysis

For the statistical analysis, R Core Team: R: A Language and Environment for Statistical Computing (R Foundation for Statistical Computing, Vienna, Austria R Version: 4 Released: 2020.04.24) was used. According to the predictor and outcome variables, different statistical tests were chosen. To investigate hand coverage, two models were investigated, one where the missed surface area that larger 0% was handled as continuous variable and the other, where surface area was handled as binary variable (0% missed and not 0% missed). In the first case, a linear mixed effect model was utilised, in which predictor values were handrub volume, type and subjective volume awareness, while the outcome variable was the logarithm of the missed surface area (%). In the second case, we utilised a generalized mixed effect model with logit link. In this model, again, the predictor values were handrub volume, type and subjective volume awareness while the outcome variable was the binary value of the missed surface area. For both cases, we adjusted for hand size and for the type of the first used handrub.

Random intercept was assigned to the different participants. Compound symmetry correlation structure for handrub amount and different power variance structure for handrub at different participants were used to fit the final model.

When examining the correlation of handrub type and volume to drying time and disinfectant spillage, a linear mixed effect model was used, where the outcome variables were the logarithm of the drying time and the fluorescence coverage, respectively.

**Table 4** Disinfectant spill test

| Question | Possible answers |
|---|---|
| Did the disinfectant spill from your hand during the rubbing? | Yes/No/I don't know |

Results are presented as mean along with ±1 standard deviation.

## Results

### Evaporation rate investigation results

According to our results, both the gel and the liquid ABHRs used in the study exhibit evaporation properties well aligned with the commercially available disinfectants. The small deviation from the other disinfectants is probably due the handrubs' different chemical composition, an important consideration for future studies investigating in vitro or in vivo evaporation rates of ABHRs. Evaporation rate depends highly on the disinfectant volume and surface area (Fig. 3, Additional file 1: Figs. S1 and S2). Overall, liquid formulations showed a similar evaporation behaviour, as after 30 s, the evaporation rate followed a linear pattern. Measurements were performed for 3 min, as, according to our previous study, application times typically did not exceed 2 min [18].

As the evaporation rates of the ABHR were similar to those, available on the market, one can safely presume that the results presented in this study would be valid for most ABHRs, although some minor differences may be observed due to the different alcohol concentration and other variations.

### Disinfectant coverage results

The participants readily understood the concept, and adhered to the measuring protocol during the assessment, however, 40 measurements were removed due to software artefacts. The final number of examined hand hygiene events was N = 641. Comparing the total missed area, liquid AHBR performed better at both volumes (Fig. 4). While at 1.5 ml, both formulations resulted in a missed area of over 5%, at 3 ml, that area decreased to 2% for the gel, and 1.1% for the liquid (Fig. 4). According to the statistical analysis, the decrease in "uncovered areas" was significant ($p < 0.001$). In addition, a significant difference was found in the missed area between students' feeling the volume was "not enough" and "just right" ($p < 0.001$). On the contrary, missed area differences were not found between feelings "too much" and "just right".

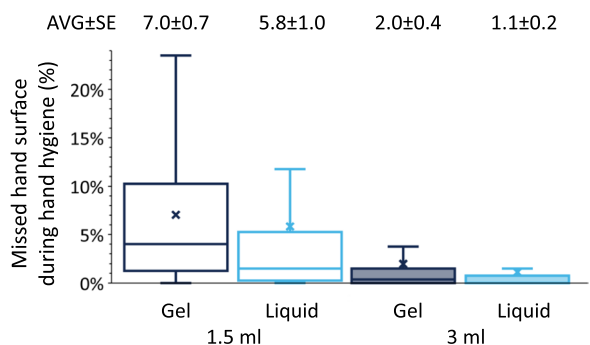

**Fig. 4** General hand coverage results (missed total hand surface area when using 1.5 and 3 ml of ABHR)

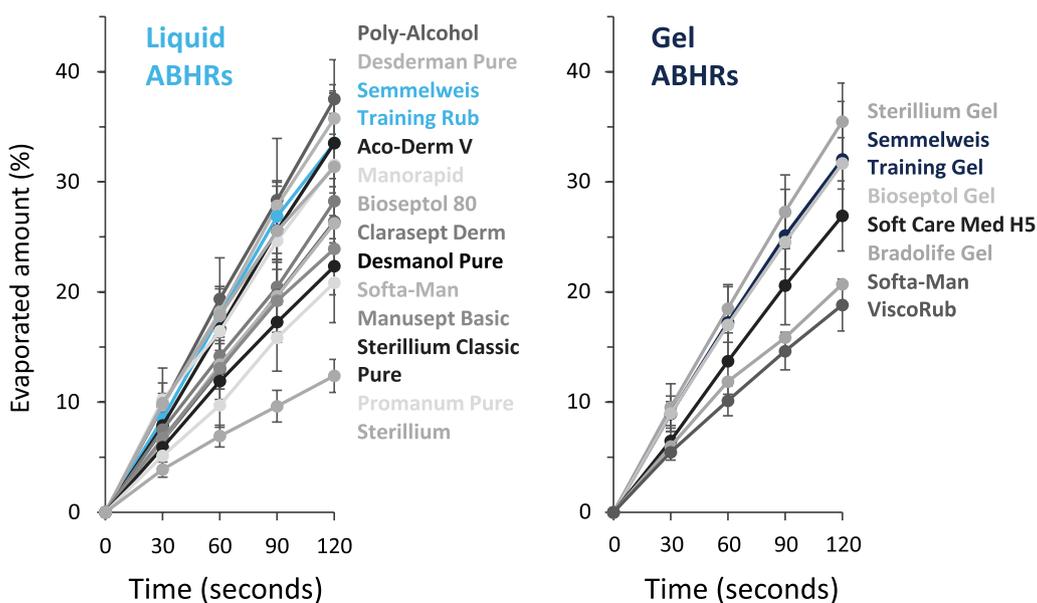

**Fig. 3** Average ABHR evaporation loss measured at every 30 s. Presented values are averaged results (Temperature: 34 °C, Surface Area: 7.07 cm$^2$)

More importantly, a significant difference was found in the missed surface area when using gel or liquid ABHR. The liquid ABHR performed better (by 1.2% and 0.9% at 1.5 and 3 ml respectively) and the difference was significant in both the linear mixed effect ($p=0.002$) and the generalized mixed effect ($p=0.004$) models. Finally, not enough statistical evidence was found to say the order of the handrubs (which formulation was used first, gel or liquid) made a difference.

Further examining the uncovered hand surfaces, the dorsum of the hands, as well as the fingertips (and especially the thumbs) were the most frequently neglected areas during hand hygiene events (Fig. 5). Interestingly, when applying the liquid handrub, the dorsum coverage was better, while no significant difference was found regarding the palm coverage.

### Hand size determination and specific hand coverage

Hand sizes were calculated for each individual, using the recorded images from the hand coverage measurements. The relative hand coverage was calculated at individual level, by dividing the applied handrub volume by the total hand surface (in $cm^2$) according to previously presented method [18]. Introducing the concept of "relative hand coverage" aims to incorporate hand sizes in the disinfectant volume–disinfectant coverage correlation. (To help to interpret the data, we mention that median hand size in this study was 355.8 $cm^2$, therefore the corresponding coverage was 2.1 $\mu l/cm^2$ for 1.5 ml or 4.2 $\mu l/cm^2$ for 3 ml of handrub). Examining Fig. 6, it is evident that increasing the volume of the gel almost linearly increased the relative coverage in the investigated volume range, while in the case of liquid, it quickly reached a plateau.

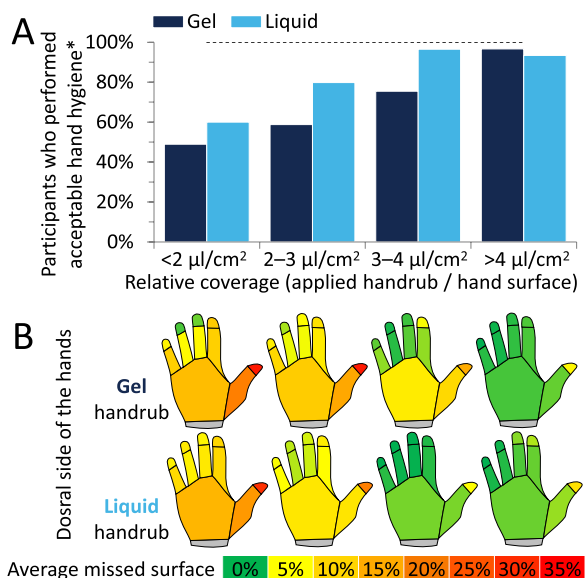

**Fig. 6** Specific hand coverage results. **A** Percentage of participants who performed acceptable hand coverage in the case of different relative coverage values *Acceptable hand coverage here means, that the participants covered at least 95% of their hands' surfaces. **B** Most frequently missed regions on the dorsal side of the hands in case of different relative coverage values

### Application time

Application length measurements showed times well above the WHO prescribed 20–30 s application time (Fig. 7). Even at 1.5 ml application volume, hand rubbing took more than 30 s (30 and 32 s in the case of gel and liquid, respectively). At 3 ml volume, application time increased further (40 and 42 s). Interestingly, the large standard deviation suggests that some other parameters induce a large variance amongst participants.

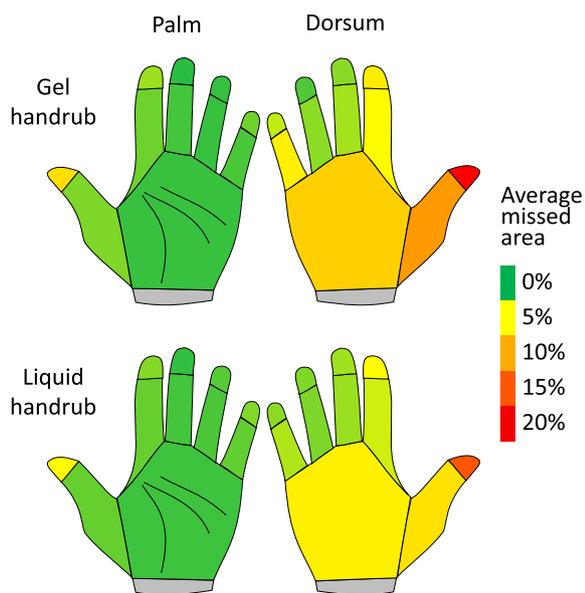

**Fig. 5** Typically missed areas during hand hygiene events. The scheme represents average values from both the right and left hands

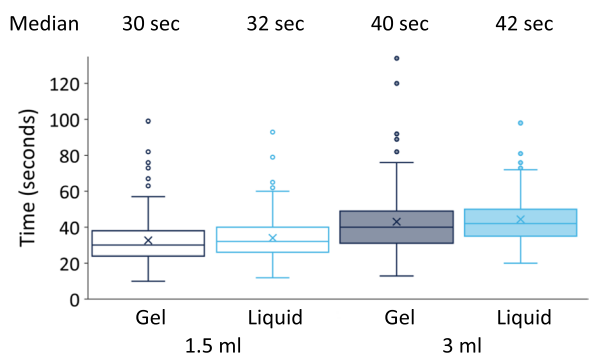

**Fig. 7** Application time results of gel and liquid formulated ABHRs

Based on the used statistical models, the difference in the drying times between using different formulation (gel or liquid) and applying different volumes (1.5 vs. 3 ml) were significant ($p = 0.007$ and $p < 0.001$, respectively; Marginal $R^2$/Conditional $R^2$:0.157/0.395). The difference between the different format is significant, however, small enough to say it has no practical significance in the clinical settings.

### Disinfectant spillage

First, comparing a subjective (Student Questionnaire) and an objective (UV Photography) method of assessing of disinfectant spillage, it is evident that although students sometimes took notice of the disinfectant's spilling during the hand rubbing event, they were not always right. The UV photography highlighted unseen spills as amount of fluorescence was detected even when students reported that they did not spill any handrub at all (Fig. 8).

Examining the hand hygiene events further reveals that at both 1.5 and 3 ml, the liquid disinfectant spilled more than the same gel volume (Fig. 9A). At increased volumes, disinfectant spillage is larger for both gel and liquid. The students truthfully reported these observations (Fig. 9B).

Incorporating hand sizes in this equation reveals that while spillage increases rapidly when using liquid ABHR, the gel spillage reaches a plateau (Fig. 10A). These spills can be critical, although they may seem negligible to the naked eye (Fig. 10B), they can be rather wasteful, as documented by the UV photography (Fig. 10C). Based on the used models, the difference in the fluorescence between using gel or liquid is significant ($109.50 \pm 8.63$ thousand pixels in case of gel and $406.19 \pm 19.88$ in the case of liquid), while the volume has a significant effect ($p < 0.001$).

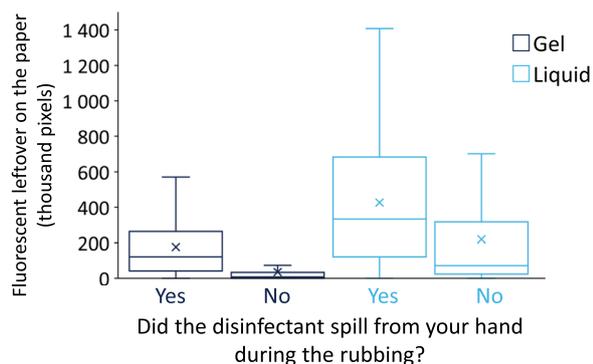

**Fig. 8** UV photography results in comparison with participants' responses regarding spillage

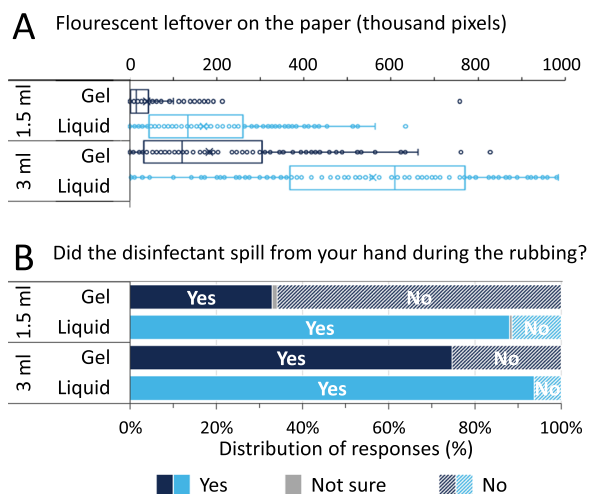

**Fig. 9** **A** Spillage at different ABHR volumes **B** and corresponding participants' responses

### Volume awareness

A basic level of volume awareness seems to be present amongst participants, as volume changes were noticed by a substantial part of them. Interestingly, at the lower volume (1.5 ml) the liquid ABHR could be felt better than the corresponding gel volume, since more students indicated it was not enough (Fig. 11A). Figure 11A shows, that 40% and 60% of the participants felt that 1.5 ml volumes were appropriate from the gel and liquid formats respectively. Additionally, 3 ml from liquid and gel felt too much for approximately 20 and 30% of the participants respectively. Coupling these results with the corresponding hand coverage results (Fig. 11B) it is evident that even when the participants felt that the given disinfectant volume was enough, approximately 15% of them missed more than 5% of their total hands' surfaces, while when feeling that the volume was too much, the same error level was approximately 6%. Important to note that the 5% threshold was chosen to present coverage difference in a simpler manner, as currently no such threshold is defined in the protocols.

### Discussion

Surprisingly, evidence-based literature regarding applicability differences amongst ABHR formulations is very limited. According to our knowledge, this is the first study examining potential coverage differences between liquid and gel ABHR formulations. Prior studies examining coverage of liquid ABHR can be found, however, unlike others and similarly to our previous work [18], a large number of hand hygiene events were documented in this study. Furthermore, hand hygiene performance

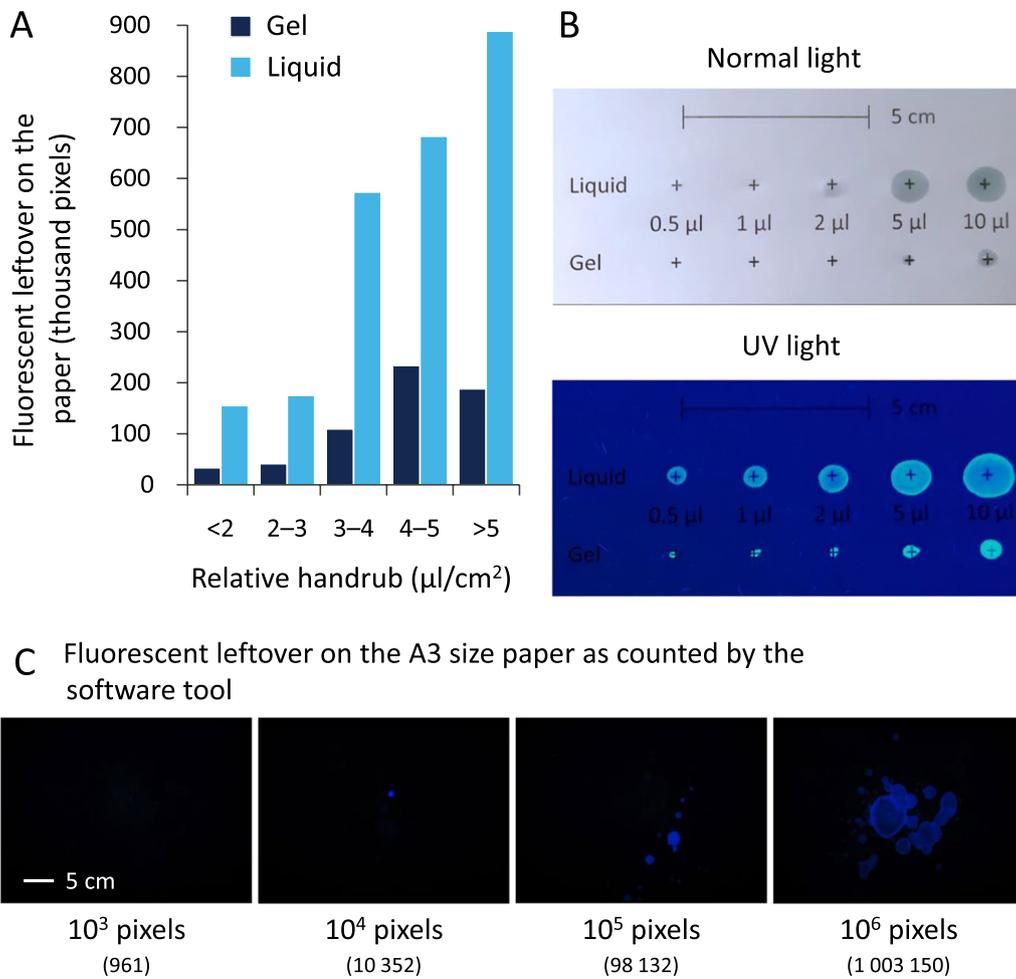

**Fig. 10** Disinfectant spillage UV photography results: **A** Relative handrub and spillage correlation, **B** Approximation of volume to intensity, **C** Representative spillage examples

was evaluated by an objective computerised and automated electronic system, and not manually by a subjective human examiner [30]. In addition, a comprehensive investigation incorporating hand sizes, volume awareness, disinfectant spillage was performed for each hand hygiene event performed during the investigation.

For being objective, before initiating this study, the dye incorporating formulations were assessed and compared to commercially available ABHRs. As the evaporation rates of the ABHR were similar to those available on the market, one can safely presume that the results presented in this study would be valid for most ABHRs although some minor differences may be observed due to the different alcohol concentration and other variations.

Volume dependent coverage results concur with our previous results, as well as studies by other researchers [16]. The 1.5 ml ABHR volume is insufficient, as in both formulations the non-covered areas exceeded the 5% of the total hand surface. In contrast, at 3 ml, the non-covered areas decreased to 2% for the gel and 1.1% for the liquid ABHR. Typically missed areas (thumbs and dorsum) were not different when using the two formulations, suggesting that the same hand hygiene technique was followed in both cases. Several previous studies also found that thumbs and fingertips are the most frequently missed areas, [31, 32], which suggests that the personal hand hygiene technique may be inappropriate in many cases.

According to our results, the difference in surface coverage between the liquid and the gel is significant, however, this is only true for the specific examination. Extrinsic (ambient temperature and humidity) and intrinsic (participants hand temperature, skin type and hydration state) parameters may present some additional limitations we faced. Nevertheless, coverage depends on the ABHR volume and on the format itself dominantly.

Investigating the relative handrub coverage, where the disinfectant volume, as well hand size is incorporated,

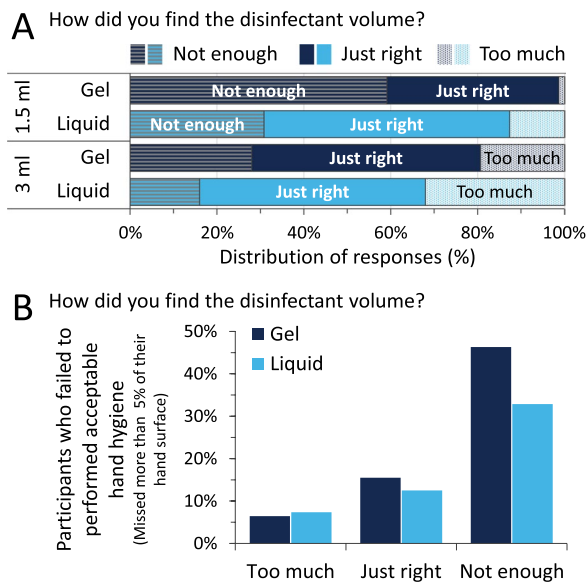

**Fig. 11** **A** Participants' responses regarding volume awareness. **B** Correlation of acceptable hand hygiene events and volume awareness

liquid and gel formulations behave differently. The liquid specific handrub coverage increases rapidly and reaches a plateau. In simple terms, while initially using a larger volume is beneficial for people with small and medium size hands, after reaching a specific value (3 µl/cm$^2$), the larger volume is beneficial only for people with larger hands as the additional volume does not only spill, but neither it increased the disinfectant coverage, similarly to the smaller or medium size hands. This phenomenon is more relevant in the case of the liquid ABHR. On the contrary, the gel-based formulation follows a steady linear increment without rapidly reaching a plateau (even after 4 µl/cm$^2$). Probably due to their viscosity, gels spill less, and therefore larger applied volumes can be better handled by HCWs, regardless of their hand sizes. This suggests that optimised technique can largely help to reach a perfect coverage. In general, the two formulations produced significantly different results when equal volumes were compared, nevertheless, additional investigations could be instrumental, if a personalised or *hand-size based volume* were examined, where formulation-based differences could potentially be present. This subsequently could result in different personalised volumes from liquid and gel based ABHRs which would optimise efficiency, increase compliance, minimise spillage and waste of resources.

The "status quo" [18] regarding application times was once again proven controversial. Even at 1.5 ml volume, application time was above the 20–30 s indicated by the WHO guidelines, regardless of formulation; whereas with a volume of 3 ml (which, according to several studies, should be required) dried in 40 s on average. Volumes that dry in under 30 s are equal or lower than 1.5 ml, by which one cannot reach complete hand coverage typically. Other works have come to the same conclusion [13, 14, 18, 19]. These contradictions should be resolved, consequently, the WHO guidelines may require a review, which process officially started last year.

Disinfectant spillage was detected by the students, however, UV proved to be an efficient tool detecting spillage events and quantifying them, even when students did not take a notice. Gel-based ABHR produced significantly smaller spills than the liquid based one. Important to note that we even used an objective digital segmentation algorithm to accurately quantify the spillage (fluorescence); liquid based ABHRs produced larger spills than anticipated, approximating even 15% of the given volumes. Another noteworthy point relates to analysing relative handrub volume and application efficacy (especially with the liquid formulation), spillage becomes even more prominent, suggesting that although larger volumes are beneficial for increasing the hand coverage of larger hands, they are wasteful at smaller hands. This could result in significant disinfectant wastes, which could in turn be costly to institutions and hospitals.

Finally, although volume awareness was present, as the difference between 1.5 and 3 ml volumes were generally noticed, it can be misleading. A large percent of the participants (25%) who deemed the disinfectant volume given to them as sufficient ("just right") produced technical errors in their hand hygiene, leading to missing more than 5% of their hands' surfaces. 6% of participants who felt the volume given to them is "too much" 6% failed to achieve at least 95% total surface coverage. In other words, participants can underestimate the volume, which they should apply. This calls for the update of teaching and training methods. Hand size was determined to be crucial, once again. For example, even at 1.5 ml, 7% of students felt the volume as "too much", while even designating the volume given to them as insufficient ("not enough") some participants failed to reach a complete coverage of > 95%. Therefore, it can be stated that optimal ABHR volume is relative. 3 ml can be considered as good enough volume in average terms (uncovered areas < 1.5%). Personalised volumes could provide an overall solution to this issue. By giving HCWs a volume calibrated to their hand can have three positive effects: coverage will increase for larger handed people, spillage and material waste will decrease, and compliance might increase as well since reports described that less rubbing time is preferred by HCWs. This is in line with the recent finding on the influencing factors of hand hygiene learning curves, measured in a clinical setup [33].

## Conclusion
The format of the ABHR was not proved to be crucial. Formulation significantly affected the coverage, although the difference in coverage was smaller than when two different volumes were compared. Liquid ABHR produced significantly more spillage than the gel-based ABHR.

Volume of the ABHR proved to have a very important role; 1.5 ml was insufficient to properly cover the hand surface typically. Increasing the application volume of a liquid ABHR (from 1.5 ml to 3 ml) was not always beneficial, as it resulted in significantly more spillage, while it did not increase hand surface coverage. Drying times (for both formats and both volumes) were far greater than the suggested application times. We were not able to find a volume that would be universally suitable for everyone. The implementation of a personalised or hand size-based volume can help to balance between maximising the coverage and minimizing the spillage and drying time. Furthermore, results suggest that the optimized volumes could be different in the case liquid and gel ABHRs, as increased volume of gel-based ABHR could be better handled thus enhanced the coverage without as much spillage.

In conclusion we can say that based on the parameters we measured during our study, we have not been able to prove that one format would be inferior to the other.

### Abbreviations
ABHR   Alcohol-based handrub
HCWs   Healthcare workers
WHO    World Health Organization

## Supplementary Information
The online version contains supplementary material available at https://doi.org/10.1186/s13756-023-01212-4.

> **Additional file 1: Fig. S1.** Liquid ABHR—in vitro evaporation rate examination results. **Fig. S2.** Gel ABHR—in vitro evaporation rate examination results.

### Acknowledgements
The authors would like to thank to all the participants who contributed to the study. We would also like to thank the HandInScan Zrt. to provide us the Semmelweis System and for their professional support during the study. Figures were designed with Biorender.

### Author contributions
CV, SB, and TH contributed to the study design. Data collection was carried out by CV and SB. Data analysis was carried out by SB, PS and DSV. All authors contributed to writing, editing, and approving the manuscript. Project supervision and financial support was provided by AJH, AS and TH. All authors read and approved the final version of the manuscript.

### Funding
Open access funding provided by Semmelweis University. Supported by the ÚNKP-21-4-I-SE-33 for CV and ÚNKP-21-4-I-SE-23 for SB new national excellence program of the ministry for culture and innovation from the source of the national research, development and innovation fund. This project has been partially supported by the National Research, Development, and Innovation Fund of Hungary, financed under the TKP2021-NKTA-36 funding scheme (Development and evaluation of innovative and digital health technologies—Evaluation of digital medical devices: efficacy, safety and social utility).

### Availability of data and materials
The datasets used and/or analysed during the current study are available from the corresponding author, based on a reasonable request.

## Declarations

### Ethics approval and consent to participate
Not applicable.

### Consent for publication
Not applicable.

### Competing interests
HT is founder of HandInScan Zrt., developer of the Semmelweis System. SB is an employee of HandInScan Zrt.

### Author details
[1]Department of Surgery, Transplantation and Gastroenterology, Semmelweis University, Budapest, Hungary. [2]Laboratory of Nanochemistry, Department of Biophysics and Radiation Biology, Semmelweis University, Budapest, Hungary. [3]Doctoral School of Health Sciences, Semmelweis University, Budapest, Hungary. [4]Department of Biophysics and Radiation Biology, Semmelweis University, Budapest, Hungary. [5]University Research and Innovation Centre (EKIK), Óbuda University, Budapest, Hungary. [6]Austrian Center for Medical Innovation and Technology (ACMIT), Wiener Neustadt, Austria.

## Publisher's Note

Springer Nature remains neutral with regard to jurisdictional claims in published maps and institutional affiliations.